\documentclass[]{nrel} 

\usepackage{verbatim}
\usepackage{wrapfig,lipsum,booktabs}
\usepackage[english]{babel}

\usepackage{amsmath}
\usepackage{graphicx}
\usepackage{titlesec}
\usepackage{float}
\usepackage{siunitx}
\usepackage[inkscapeformat=png]{svg}

\usepackage{todonotes}
\usepackage[many]{tcolorbox}    	

\definecolor{main}{HTML}{5989cf}    
\definecolor{sub}{HTML}{cde4ff}     
\definecolor{main2}{HTML}{D4AC0D}    
\definecolor{sub2}{HTML}{F9E79F}     
\definecolor{sub3}{HTML}{C1E1C1}    
\definecolor{main3}{HTML}{478778}     
\definecolor{sub4}{HTML}{DDA0DD}    
\definecolor{main4}{HTML}{800080}     

\tcbset{
    sharp corners,
    colback = white,
    before skip = 0.2cm,    
    after skip = 0.5cm      
}                           

\newtcolorbox{tipbox}{
    sharpish corners, 
    colback = sub, 
    colframe = main, 
    boxrule = 0pt, 
    toprule = 4.5pt, 
    enhanced,
    fuzzy shadow = {0pt}{-2pt}{-0.5pt}{0.5pt}{black!35} 
}

\newtcolorbox{scode}{
    sharpish corners, 
    colback = sub2, 
    colframe = main2, 
    boxrule = 0pt, 
    toprule = 4.5pt, 
    enhanced,
    fuzzy shadow = {0pt}{-2pt}{-0.5pt}{0.5pt}{black!35} 
}

\newtcolorbox{sjob}{
    sharpish corners, 
    colback = sub3, 
    colframe = main3, 
    boxrule = 0pt, 
    toprule = 4.5pt, 
    enhanced,
    fuzzy shadow = {0pt}{-2pt}{-0.5pt}{0.5pt}{black!35} 
}

\newtcolorbox{ssystem}{
    sharpish corners, 
    colback = sub4, 
    colframe = main4, 
    boxrule = 0pt, 
    toprule = 4.5pt, 
    enhanced,
    fuzzy shadow = {0pt}{-2pt}{-0.5pt}{0.5pt}{black!35} 
}
\newcommand*\iftodonotes{\if@todonotes@disabled\expandafter\@secondoftwo\else\expandafter\@firstoftwo\fi}  

\makeatother

\title{A Beginner’s Guide to Power and Energy Measurement and Estimation for Computing and Machine Learning}

\author[1,*]{Akshaya Jagannadharao}
\author[1]{Nicole Beckage}
\author[1]{Sovan Biswas}
\author[2]{Hilary Egan}
\author[2,3]{Jamil Gafur}
\author[1]{Thijs Metsch}
\author[1]{Dawn Nafus}
\author[1]{Giuseppe Raffa} 
\author[2,*]{Charles Tripp}
\affil[1]{Intel Corporation}
\affil[2]{National Renewable Energy Laboratory}
\affil[3]{University of Iowa}
\affil[*]{Corresponding Authors}

\fancypagestyle{plain}{}
\begin{document}

\frontmatter
\footnote{The first author provided the initial complete draft of this work. All other authors contributed equally. The Intel authors would also like to thank Shachi Kumar, Felix Leung, John Browne, Lama Nachman, Shahid Sheikh, Lowren Lawson, Deborah Bernstein, Eric Dahlen, and Emanuel Moss for their invaluable inputs. Any errors are strictly the responsibility of the authors. The NREL authors would like to thank David Sickinger and Kevin Griffin for their invaluable contributions. This work was authored in part by the National Renewable Energy Laboratory, operated by Alliance for Sustainable Energy, LLC, for the U.S. Department of Energy (DOE) under Contract No. DE-AC36-08GO28308. The views expressed in the article do not necessarily represent the views of the DOE or the U.S. Government. The U.S. Government retains and the publisher, by accepting the article for publication, acknowledges that the U.S. Government retains a nonexclusive, paid-up, irrevocable, worldwide license to publish or reproduce the published form of this work, or allow others to do so, for U.S. Government purposes. }




\clearpage
\tableofcontents
\newpage
\listoffigures
\newpage
\listoftables

\mainmatter
\pagestyle{fancy}


\textbf{Abstract}

Concerns about the environmental footprint of machine learning are increasing. While studies of energy use and emissions of ML models are a growing subfield, most ML researchers and developers still do not incorporate energy measurement as part of their work practices. While measuring energy is a crucial step towards reducing carbon footprint, it is also not straightforward. This paper introduces the main considerations necessary for making sound use of energy measurement tools and interpreting energy estimates, including the use of at-the-wall versus on-device measurements, sampling strategies and best practices, common sources of error, and proxy measures. It also contains practical tips and real-world scenarios that illustrate how these considerations come into play. It concludes with a call to action for improving the state of the art of measurement methods and standards for facilitating robust comparisons between diverse hardware and software environments.

\chapter{Introduction}

    The impacts of climate change are increasingly evident; the past decade has been reported as the warmest on record \citep{hansen2024global}, with new global temperature records now being set  monthly \citep{younger2024}. Meanwhile, the International Energy Agency (IEA) expects energy consumption by data centers to double by 2026, with Artificial Intelligence (AI) being a major contributor \citep{IEA2024}. Some localities are keeping fossil fuel sources online that were planned to be shut down due to this new demand for compute \citep{Halper2024,Perkins2024}. In response, the Machine Learning (ML) research community has started recognizing its growing energy use and contribution to climate change. 
    
    ML research has explored energy use from various angles. Initial studies focused on training \citep{strubell2020energy}, later expanding to inference costs \citep{luccioni2024,DESISLAVOV2023100857} and energy trade-offs throughout the development lifecycle \citep{wang-etal-2023-energy}. For a comprehensive review, see \citet{verdecchia2023systematicreviewgreenai}. Outside of ML, data center researchers have also looked at energy use to improve grid stability alongside emissions reductions \citep{lin_adapting_2023}. This growing subfield relies on accurate energy measurement of specific computing tasks. 
    
    However, even a superficial look reveals differences in measurement approaches. For example, \citet{luccioni2024} use the open-source package Code Carbon, while \citet{wang-etal-2023-energy} use both Code Carbon and a physical energy meter. \citet{DESISLAVOV2023100857} base calculations on published Thermal Design Power (TDP) of hardware. \citet{bannour:hal-03435068} note further discrepancies in how measurement tools calculate power, what hardware they account for, and how they handle Power Usage Effectiveness (PUE) in data centers. \citet{Grant2017} discuss the pros and cons of using physical meters versus software counters, but this is just one of many challenges developers face. 
    
    Given the widespread concerns about the sustainability of ML and software in general,  the ability to measure energy consumption is crucial. Energy, in theory, is measurable as an objective property of physics. There is a physical link from the electricity source to the software. However, in nearly all practical scenarios, a surprising amount of judgment and knowledge is required to measure this link accurately. Modern computing hardware and software are highly complex systems, where different settings, configurations, and conditions can mean that running the same workload on seemingly "the same" system can give different results. There are no fully standardized tools and benchmarks that work across all situations, and estimates can vary in usefulness.  To give some examples, energy usage of specific hardware can be measured based on its actual power draw, but correlating these measurements to an underlying workload, or part of a workload, can be challenging. Meanwhile, measurement closer to the point of software performance can be straightforwardly correlated to the workload; however, this may not capture the full impact of the workload on the system. Software based measurement can also introduce additional energy utilization overhead. Finally, the interaction between software and hardware can introduce further variability, as different hardware architectures and configurations can affect how software performs and, consequently, how much energy it consumes. For those wanting to go further and map energy on to carbon emission, yet more accounting complications lie ahead (see Section 3).  
    
    Making informed decisions about a measurement approach is vital. This article serves as an introduction to energy estimation for those needing to estimate energy use for the first time, whether they are ML developers or other software researchers and practitioners. We aim to provide a comprehensive overview of the principles and methods involved in estimating energy consumption, offering practical guidance on how to approach this task. We discuss the measurement of energy usage from both hardware and software perspectives as well as the challenges of interpreting these measurements as useful estimates.   By understanding the factors that influence energy usage and the techniques available for measurement and estimation, readers can gain insights into optimizing both hardware and software to reduce energy consumption. This knowledge is increasingly important in an era where energy efficiency and sustainability are critical considerations in the development and deployment of computing technologies.

\chapter{General Approach to Power Measurement Analysis}
\label{general_approach}
In this section, we outline a generalized methodology for power measurement and analysis. We first introduce a simple and  generic way to approach energy measurement problems (Section \ref{sec:workflow}), followed by a few commonly occurring situations where the need to measure arises (Section \ref{sec:scenarios}). Section \ref{sec:arch} then contains a high-level description of an abstract computing facility for reference (Section \ref{sec:arch}).  From there we transition to a more practical discussion of the choices associated with various aspects of energetic measurement (Section \ref{sec:challenges}). We then discuss proxy measurements, which become relevant when direct measurements are either unavailable or challenging to obtain (Section \ref{sec:proxies}). We conclude the section with a discussion on the analysis of these measurements and how they can be correlated with the quantities of interest (Section \ref{sec:interpretation}).

\section{A Basic Workflow}
\label{sec:workflow}
\begin{figure}[h!]
    \centering
    \includegraphics[width=.8\linewidth]{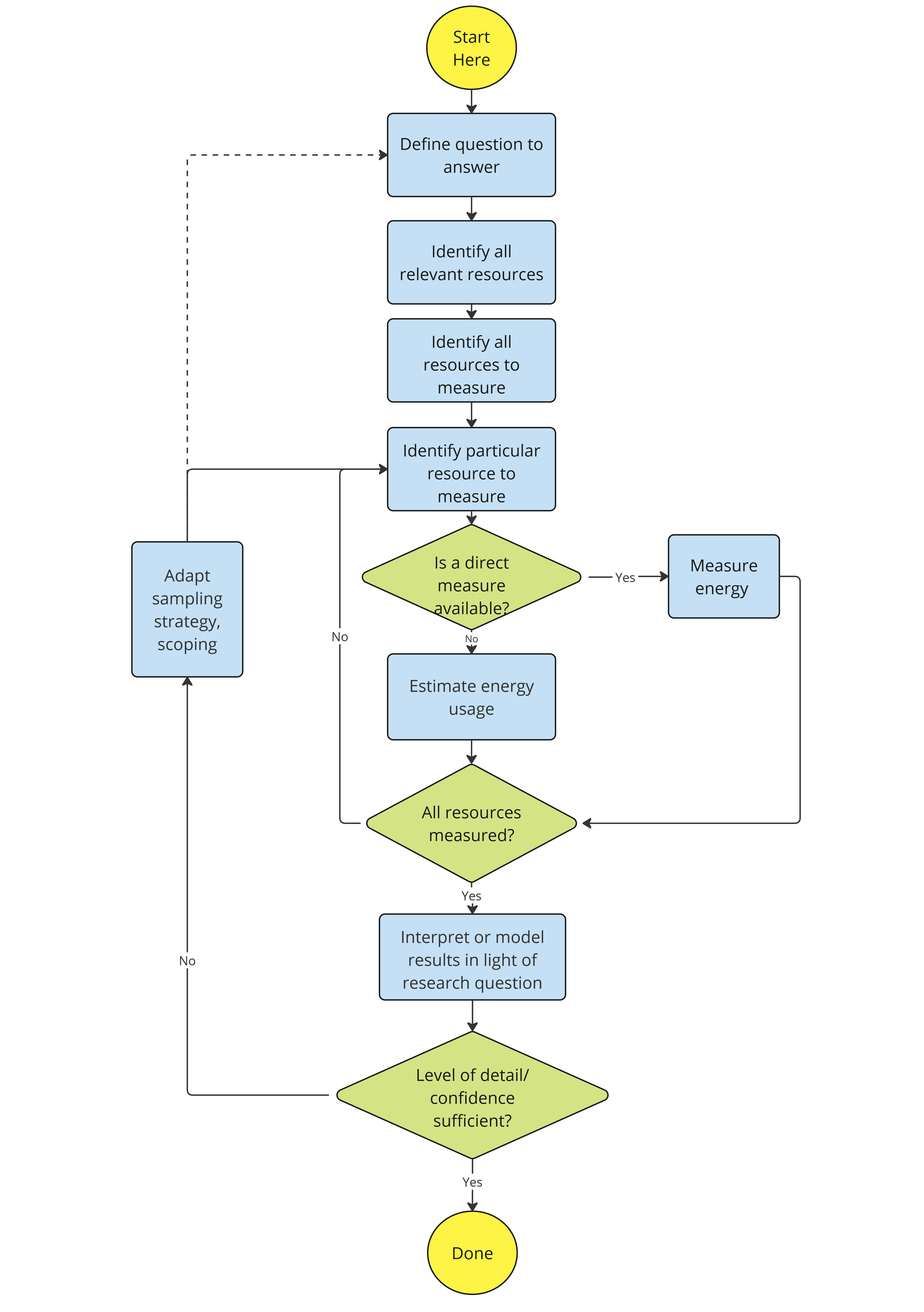}
    \caption{Steps for an Energy Measurement Project}
    \label{fig:flowchart}
\end{figure}

Energy measurement starts with the definition of the problem or question, and then involves a series of decisions regarding where and what to measure - and how often- in an often iterative process.  To describe the steps and types of iterations one can expect, we offer a flowchart (see Figure \ref{fig:flowchart}) based on the utilization, saturation, and errors (USE) method of performance analysis (see \citet{GreggUSE}).

As shown in Figure 1, broadly speaking, the first step is to determine the questions you wish to answer, and then identify what resources are relevant to the problem you are trying to solve. Next, clarify any difference between what is relevant and what is in scope for measuring. Some relevant resources might be useful to know about, but not useful to measure as part of the project. For example, if a developer has no choice in what data center they use, overall data center efficiency might not be relevant, but if they do have a choice to work in a more efficient data center, it might still be relevant information even if it is not being measured as part of the project. Assessing whether larger efficiency gains can be obtained via modification of the software or the system itself (or co-optimization of both) is key part of many research questions that should be explicitly formulated, even if your project ultimately focuses on one side of the equation or the other.

The next step is to determine what measurements are feasible and appropriate for each resource. \emph{Direct measures} are typically taken either by a physical meter or through telemetry reported by the hardware. However, in cases where direct measurement is not feasible, \emph{estimates} that rely on a combination of direct and indirect power measurements, and/or proxy measurements, may be available  (see Section \ref{sec:proxies}).  When you are defining where and how to measure or estimate, drawing a block diagram can help clarify if you have the right scope. At that point, you also begin to assess whether that particular measure, at that sampling rate, is sufficient. Next (and, more realistically, in parallel), you turn your attention to the next resource that needs measuring, until all the resources in scope are accounted for. Expect to take measurements multiple times for each resource, as this paints a more reliable picture.    

Once you have data, you then proceed to the analysis and interpretation to turn that data into insight into the problem at hand. Most situations involve interpreting the data relative to some amount of work done-- an image generated, a feature executed. Often that can be relative to a previous version of the software. At this stage it is important to explicitly fix all other aspects of the problem to create a valid comparison or (where not possible) statistically average over non-fixed aspects (for a full discussion see Sec \ref{sec:challenges-stats}).

 Finally, the interpretation stage involves evaluating the sufficiency of the analysis itself: is the data convincing? Do we have all the details we need? If not, you might need to adjust the measuring strategy, and on rare occasions even change the question to something for which evidence can be collected. It is also possible to go \textit{too} deep. Do get as detailed as you feel confident, but should your find yourself twisted around in one nuance or another, revisit your primary goal and re-establish what is truly mission-critical.

\section{Common Types of Measuring Goals}
\label{sec:scenarios}
\begin{figure}[hbt!]
    \centering
    \includegraphics[width=.8\linewidth]{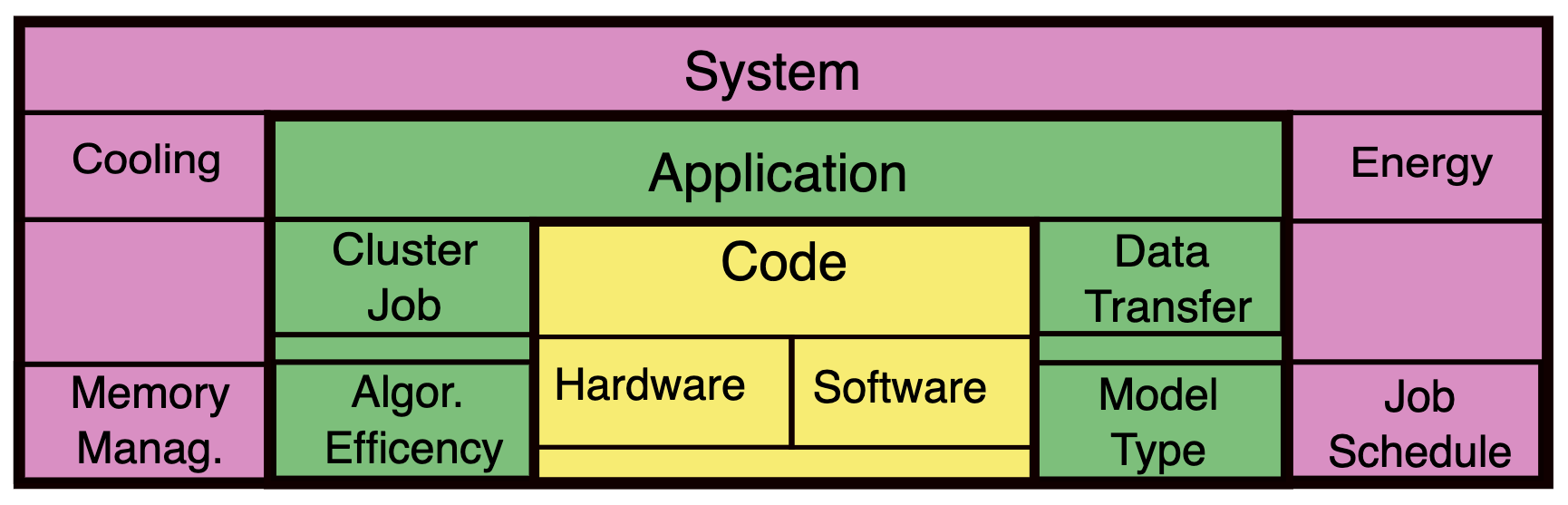}
    \caption{The figure illustrates the progression from specific to broader scopes of energy measurement in computational contexts. Each color represents a distinct measurement scenario, connected in a logical sequence.}
    \label{fig:Example_outline}
\end{figure}

Measurement goals, or research question, we can coarsely categorize into three levels: system, job or application, and code. Figure \ref{fig:Example_outline} illustrates some resources that are often considered for measurement when asking questions at these different levels.  

\subsection{System Level}
\label{sec:sys}
Here we define a system-level scenario as assessing the total energy used by a collection of workloads holistically across an entire system.  Example use cases include estimating the overall computing energy used at a company level, investigating how different policies impact total energy consumption at a data center, or assessing the optimal spatio-temporal placement of \emph{collections} of jobs (distinct computing workflows) across varying resources. "System" here is defined by the research intent, not by equipment types.  Carefully bounding the system in question is key, which ensures measurements can be attributed appropriately to the unit of analysis, whether it is a workstation, data center, or corporate division. In this scenario, the impact of individual jobs or applications is typically not dissaggregated, and analysis instead focuses on capturing the overall effect of their interactions within the system. At the system level, broad measurements can provide insights into overall energy consumption and efficiency. However, per-node and per-job instrumentation can be valuable for understanding the causal relationships between workloads and total system energy draw. 

\begin{ssystem}
    \textbf{Real World Example (S1):} At the National Renewable Energy Laboratory (NREL), data center researchers investigated the impact of job scheduling policies on power consumption to avoid peak power pricing \citep{bugbee2017prediction}. 
\end{ssystem}

\subsection{Job/Application Level}

    Measurements at the job/application level address concerns that lie between code-level and system-level measurements. This perspective is concerned with questions like, ``How can I adjust or configure my jobs to be more energy-efficient?'' or estimating the total energetic cost of a particular workflow. In this scenario the background workloads of other jobs on a system are ignored/held ``fixed'' and care must be taken to ensure statistical validity of such comparisons. At the job level, per-node or per-job measurements are almost essential for drawing meaningful conclusions. Without these measurements, the results can be confounded by interference from other system activities. Scenarios of this type may involve comparing the same workload across different hardware, comparing different software approaches to an overall computing task, or co-optimization of both.
    

\begin{sjob}
    \textbf{Real World Example (S2):} At NREL we explored the energy impacts of varying neural network architectures and hyperparameter selections across multiple training workloads~\autocite{ButterEnergyPaper,butter_e_dataset}, we estimated the total energy consumption per job on two base hardware configurations: CPU nodes and GPU nodes. We standardized the energy estimates across the jobs in order to remove the effect of different hardware configurations. 
\end{sjob}

\subsection{Code Level}
\label{sec:code}
Code-level measurements focus on quantifying the power used by specific parts of a larger job or application. Examples include estimating the energy consumption of a single epoch in a deep neural network architecture or explorations into how to pre-process data before sending it to the GPU to minimize energy use. This type of measurement can provide insight into which parts of a larger job consumes the most energy and allows for comparing the efficiency of different algorithmic implementations. Code-level measurements are often used as a preliminary step towards creating an energy profile for a segment of a larger project. Code-level measurements generally require high temporal resolution and are typically more concerned with relative comparisons of energy consumption rather than absolute measurements. Tools that estimate energy consumption based on hardware counters and system utilization metrics can provide reasonable relative estimates for many use cases. However, software-based measurement metrics can interfere with the performance and energy profile of the code being tested. Additionally, attributing energy consumption to specific lines of code can be particularly challenging, especially in multithreaded, multitasking, and multitenant environments.

\begin{scode}
    \textbf{Real World Example (S3):} At Intel Labs, we explored how to minimize the energy impact of a specific neural network architecture during both training and inference. We found that different deep neural network architectures have different points where energy consumption can be optimized and power usage decreased. This further interacted with the hardware configuration that the code was running on.

\end{scode}

Gaining clarity on scope of the research question, and scope of the measurements, is important but can also be difficult because resources are shared across these levels. Asking what the impact could be of hardware upgrades across an entire data center is different from asking about the energy implications of training a graph neural network (GNN) on that same hardware. Both require energy measurements from the same hardware, but at the data center level, you are more likely to be interested in that hardware's power draw as a percent of all the other resources--cooling, memory management, networking, etc.,-- whereas in the code level, your central concern might be in making sure you have a clear understanding of how energy draw maps on to the different phases of activity within the code. 

\section{Architecture of a System}
\label{sec:arch}
The previous section focused on defining the kinds of problems you might want to solve, how those relate to the scope of measurement. Where you \textit{take} the measurement, however, breaks down someowhat differently from the levels outlined above. The first measurement decision starts with identifying the specific locations and scale at which measurements will be taken. Here we outline a generalized hierarchy of an abstract system for reference throughout the rest of the paper, in alignment with recommendations from the Energy Efficient Computing Working Group Procurement Guidelines \citep{eehpc_procurement}.

\begin{itemize}

    \item \textbf{Data Center/Facility}: At this broadest level, a data center or facility measurement comprises the entire compute and supporting infrastructure. There may be multiple systems in a given facility (e.g. different HPC systems within a single building). 

    \item \textbf{System:} The system level may vary by site and architecture, but includes all of the parts of a single system that explicitly participate in performing any workload(s). This might include supporting internal and external power and cooling equipment as well as internal and external communication and storage subsystems.    

    \item \textbf{Platform:} The platform is distinguished from the system so as to differentiate compute from other subsystem equipment (such as external storage) that may be managed distinctly, but together comprise a system.

    \item \textbf{Cabinet/Rack: } The cabinet (or rack) is the first order discretization of the system measurement. The cabinet may be part of the compute, storage or networking platform. A cabinet or rack is made up of multiple nodes.

    \item \textbf{Node:}  A node consists of the combination of all components that make up a discrete compute unit for the architecture. For example, components may include the CPU, memory and the network interface. The node may be part of the network or storage equipment, such as network switches, disk shelves, and disk controllers.

    \item \textbf{Component/Device Level}: Components are the physically discrete units that comprise the node. Components can be any devices that are part of a node for a particular architecture (e.g., CPU, GPU, RAM, NIC, SSD). 
\end{itemize}

Those working in situations closer to Scenario 1 (S1, above) will be more likely to be interested in data center and system levels, and those with code level questions are more likely to have interest at the component level, but it might not always perfectly break down this way. For example, you might be interested in code-level optimizations, but when deciding where to run a job, any information that a data center owner can provide about overall efficiency, or component efficiency, can be useful information. Measurements at each of these levels can be taken in multiple ways and have their own challenges and advantages, as outlined in the following section.
\section{Common Challenges and Considerations in Energy Measurement}
\label{sec:challenges}
Here we outline a non-exhaustive list of the challenges that can frequently occur when measuring or estimating energy use in computing scenarios such as those outlined in Section (\ref{sec:scenarios}), specifically focusing on those that a software developer or similar user may encounter. Where possible we also provide generic recommendations for approaching said challenges and specific examples of real-world test cases.

\subsection{Idle Power Draw}
Turning on a computing system requires energy, regardless of if the system is being utilized for a computing workload. Therefore is is necessary to be precise about whether one is measuring the additional energy (over baseline) it requires to run a workload or the total energy consumed by the system during the time period it runs a workload. This trade-off is closely related to the idea of marginal vs absolute energy estimate. In an \textbf{absolute estimate} one takes into account all power consumed (even by background processes), whereas a \textbf{marginal estimate} measures the change in energy consumed relative to a baseline condition.

Absolute vs. marginal estimates allow for different understandings about the energy consumed and the effect of hardware on software. For example, the marginal energy consumed over that of an idle system bounds the possible reduction in power, however, this requires very careful understanding of the baseline energy usage which can be highly variable. Different nodes or components of the same architecture may have different idle power draws due to small variations in the hardware or previous usage patterns. In contrast, absolute energy is easier to measure, but comparing between different workloads can be misleading if the underlying basline changes.

\begin{tipbox}

\textbf{Tip!}
Where possible, report both absolute energy use and marginal power use for a full picture of effective resource usage.
\end{tipbox}

Comparing only marginal energy use can also lead to misleading conclusions. Imagine a scenario when one is optimizing a code that can effectively use 2 processors to full performance, and comparing the energy use of running this code on Device A with 2 cores vs Device B with 36 cores, where Device A is a slightly slower chip with a lower total baseline power draw but similar non-idle use per core. If one only compares marginal energy use it may appear advantageous to run the workflow on Device B as it will complete slightly faster and therefore lead to a lower marginal energy estimate; however, the absolute energy cost will be greater due to hardware over-provisioning.

\subsection{Supporting Infrastructure/Cooling}
The difference between an absolute vs marginal energy estimate is a particularly important distinction when deciding whether or not to include measurement of associated infrastructure beyond compute resources such as cooling systems or storage. While running a more power intensive computing workload will typically require a subsequent increase in the amount of cooling power, this relationship is both non-linear and typically involves a time-lag as cooling is ramped up and down. 

These factors make it difficult to directly associate the contribution of the supporting infrastructure to an energetic estimate  with a particular pieces of code; therefore, when looking to optimize a particular piece of code (e.g. S3) this contribution is typically ignored.  

In contrast, when estimating the total energetic impact of a particular workload (e.g. S2) particularly across different systems the contribution of supporting infrastructure may be important. On an isolated system  one may sum the impacts of the energy usage of these resources as well; however, care must be taken to establish effective baselines for these resources as well. For instance, there is often a strong weather dependence in the amount of energy cooling takes which must be accounted for in any direct comparison.

One generic way to estimate the contribution of cooling infrastructure is to scale the energy measured by the typical Power Usage Effectiveness (PUE) of the system. Systems with a lower PUE (values closer to 1) use a lower fraction of their total power usage on cooling compared to power used for compute resources, and are therefore more efficient for fixed computing infrastructure. While this is an imperfect measure on its own (see further discussion of other metrics such as Total Power Usage Effectiveness (TUE) and IT-power Usage Effectiveness (ITUE) \citep{patterson2013tue}), it is a commonly used first step when comparing various resource options for workload deployment. 

\subsection{Measurements on Shared Resources}
A similar, but even more difficult to disentangle challenge arises when performing energy measurements on shared resources like HPC systems or cloud data centers. Processes that are unrelated to the workload or code of interest occupy the same shared resources, be it shared compute in a multi-tenant system or shared network/storage/cooling infrastructure. Disentangling what proportion of the shared resource usage should be attributed to a particular workload in either a marginal or absolute energy measurement can be extremely difficult, and is therefore typically neglected. However, co-optimization of hardware and software may require assessing contributions from these aspects as well, particularly for large-scale workflows; this is an open research challenge.

Furthermore, other processes may directly impact the performance of the workload or code of interest; for instance, a busy network can slow down inter-node communication, causing more CPU cycles to be spent idling. Software profilers and close collaboration with facility experts can help researchers assess the impact of e.g. degraded internode fabric performance due to high volume network traffic on the workflow and associated energy. In either case, care should be taken to perform multiple experiments to average out the impacts of other background workflows on shared resources.

\begin{tipbox}

\textbf{Tip!}
For best results, low-level code optimization is best performed on an isolated system when possible. In all scenarios where shared or background processes \emph{are} present, appropriate statistical care should be taken to average over system/background variation impact on the workload of interest.
\end{tipbox}

\subsection{In-Band vs Out-of-Band Measurements}
Measurement often happens either through built-in hardware performance monitoring tools or at the wall (also called "in-band" and "out-of-band" in \cite{Grant2017}, or on-device and off-device measurement). 
In-band measurements depend on software solutions such as those listed in the tools section (Appendix 3) to access performance monitors. For many software measurements, values reported are estimates or approximations of the power consumed by that component.

In contrast, at-the-wall measurement refers to the practice of measuring the total power consumption of a system at the point where it connects to the power outlet. This requires installing a power meter directly between the system and the wall and combining these power measurements with the workloads and running time. Power meters must also be carefully calibrated to avoid misleading results. Measuring power consumption at the wall accounts for inefficiencies in power conversion, distribution losses, and environmental factors (e.g., temperature, humidity) that can affect overall energy consumption. While this gives the most accurate measurements that ensure all components and inefficiencies are being captured, this approach typically measures power related to the entire system or rack which can include background or shared processes and are therefore more difficult to associate with a particular workload. 




\subsection{Measurement Frequency}
Coordinating sampling intervals across multiple measurement points or components is essential for synchronized data collection and accurate comparative analysis. Inconsistent sampling intervals can lead to discrepancies in the measured data and compromise the reliability of the results. Therefore, it is vital to understand the variation within the tasks of interest--seasonal variation of jobs in a data center, bursts of activity within an ML training run, etc.-- and ensure the sampling method comprehends these. 

Sampling inherently involves a trade-off between data granularity and measurement precision. Finding the optimal sampling frequency involves considering the application's specific requirements, including the desired level of detail in the power consumption data and the acceptable level of measurement uncertainty. Higher sampling frequencies result in finer granularity, providing more data points and capturing rapid changes or fluctuations in power consumption with greater detail. However, excessive sampling can introduce error due to the computational burden on the components being measured, obscuring the quantities of interest (we refer to this problem as \emph{sampling resource overhead}). This problem may be mitigated with off-device measurement techniques, but the need for off-device measurement techniques can often be avoided by sampling at the lowest frequency to answer the question of interest reliably.
\begin{tipbox}
\textbf{Tip!}
Pay attention to the relative duration of the job, process, algorithm, function, etc. being measured. If the duration being measured only allows for a small number of samples, multiple repetitions may be needed to accurately assess of energy consumption.
\end{tipbox}

\begin{tipbox}
\textbf{Tip!}
Instantaneous measurements must be carefully interpreted --- they can miss transient spikes and dips, or hit them just right giving an inaccurate picture of the overall energy usage when integrated over time.
\end{tipbox}

Software-based power reporting mechanisms can have known issues that need to be considered to prevent errors from creeping into the calculations. When looking at the collected data, we suggest checking that the timestamps align in expected ways, and that the granularity that results from the instrumentation frequency is uniform. Data might need to be cleaned or resampled as a result.  For the tools that measure based on statistical sampling of clock interrupts, like top, additional precaution is necessary if the task being measured is primarily clock driven. In addition, tools may have sampling restrictions, like the python package psutil, which says that sampling too frequently can lead to unreliable readings. Their suggestion is to sample greater than 0.1 seconds.

\begin{scode}
    \textbf{Real world learning (S3):}
     To understand the energy consumed during a single epoch of a neural network training, we estimated CPU and GPU power through independent measurements. While the power estimation measurements had the same sampling rate and were called in succession, we rarely, but not never, saw that the power estimation was in different functions within the larger code. 
\end{scode}

\begin{table}[H]
    \centering
    \begin{tabular}{p{0.23\linewidth}|p{0.72\linewidth}}
        \textbf{Sampling Frequency} & \textbf{Example Scenario} \\\hline
        High-frequency & Real-time monitoring \& response of IoT device power consumption for system anomalies\\\hline
        
        Low-frequency & Battery life estimation for mobile devices; excessive sampling may drain the battery faster due to sampling resource overhead leading to inaccurate estimations of battery life and reducing the reliability of the measurement. 
    \end{tabular}
    \caption{Examples of contexts for high and low frequency sampling}
    \label{tab:sampling_frequency}
\end{table}

\subsection{Component Interaction}
Your application might use heterogeneous resources including CPUs, GPUs, networking cards, and accelerators, each with its own unique power characteristics and monitoring capabilities. This diversity makes obtaining consistent and comprehensive measurements across the entire system challenging. One must ensure that the energy estimation of one resource/component does not include part of the energy estimate of another component. 
This can sometimes be very challenging as, for example, measuring power at the socket level captures some of the energy spent sending information to and from the networking cards. This issue can sometimes be avoided by choosing a consistent definition of component measurements that is used across all test comparisons, allowing for relative vs absolute estimates. However, when comparing across different system configurations, the aspects of a job that are being double counted may not be directly comparable, making relative power measurements inaccurate.

An additional aspect of considering component interactions is understanding the interaction between different components within a system. 
For example, consider a scenario where an application heavily utilizes networking resources. In such cases, network delays or congestion may lead to increased idle power consumption in other components, such as the CPU, as they wait for data. Being able to capture these types of bottlenecks requires accurately measuring power as well as understanding how the components interact. This may require more than just power measurements but additional tools to identify and detect when and why a component is being underutilized as part of a larger application/job. 

\begin{tipbox}
\textbf{Tip!}
Optimal energy effectiveness can often be attained by improving compute efficiency and limiting idle time resources and bottlenecks. Additional energy savings can come through optimization of code with respect to the memory hierarchy. 
\end{tipbox}

\begin{sjob}
\textbf{Real World Learning (S2)}
In our deep learning hyperparameter study, we found that processing stalls were a major factor in determining energy consumption, far outpacing computing energy costs.
In particular, waiting for data from higher levels of the memory hierarchy (caches, RAM, disk, network, internet) frequently left devices (CPU, GPU, RAM) drawing significant ``active idle'' power while stalled~\autocite{ButterEnergyPaper}. Such stall costs might be reduced if data transfers are initiated before previous computations are completed or by reducing active idle power draw through hardware setting optimization.
\end{sjob}

\subsection{Power Management System Settings}
Hardware specific system settings can have a key impact on both power and energy use. For instance, changing device clock speed reduces power draw. Other settings may have an impact on the power draw of a node. These settings can include – but are not limited – to effects of configuring power features of the CPU and its adjacencies (e.g. memory setup configurations). Intel® Speed Select Technology is an example for one of these power management control knobs that can be configured. It has a set of features to control the behavior of the CPU its cores and how their frequencies are managed. If evaluating differences in code and the impact on energy it is important to ensure hardware settings are fixed, but for general optimization it may be important to ensure that the hardware settings are chosen to ensure best performance for the a particular workload.

\subsection{Error and Variability}
\label{sec:challenges-stats}
There are multiple sources of variation that could lead to error when measuring power or limit the comparability of power measurements; Table \ref{tab:error_var} lists some non-exhaustive examples. To ensure you are measuring your quantity of interest rather than inherent variability in your system, you should always run multiple experiments and compare using an appropriate statistical test.

\begin{table}[H]
    \centering
    \begin{tabular}{p{0.2\linewidth}|p{0.75\linewidth}}
        \textbf{Source} & \textbf{Effects} \\\hline
        
        Temperature and environmental conditions & Temperature and environmental conditions can affect component performance and power consumption. Changes in temperature can alter electrical resistance and impact cooling systems, leading to fluctuations in power usage. Additionally, temperature variations can affect the accuracy of measurement devices themselves, compounding potential inaccuracies. \\\hline
        
        Power supply variability & Variability in the power supply voltage can affect the performance of electronic components and can lead to inaccurate readings \\\hline

        Background processes and resource sharing & In multi-tasking or multi-threaded environments, distribution of resources can impact the accuracy of measurements. Interference between tasks or threads may results in deviations from expected power values. Even if the work being measured is not multi-threaded, when the application is sharing resources with other background tasks, this adds another layer of complexity when trying to isolate work.\\\hline

        Variability in input data & This may result in the same block of code requiring more or less power which can be a source of variability in power estimates. This is especially worthy of consideration in situations like S1 where the research question is focused on comparing specific software implementation differences.\\\hline

        Machine wear and tear & Hardware degradation can affect how much power it uses to complete the same task.\\\hline

        BIOS sample rate & Some BIOS sample power measurements over a 3-5 second period and provide an average value, which could hinder attempts at higher resolution sampling.\\\hline

        Power management settings & Variation in C-/P- states caused by power management features controlled in the OS, BIOS, or elsewhere can create variation in output. Fan issues or high temperature can introduce throttling.\\\hline

    \end{tabular}
    \caption{Error and variation cause and effect}
    \label{tab:error_var}
\end{table}

    \begin{scode}
    \textbf{Real world learning (S3):}
     In measuring the energy of a single neural network training epoch, we saw a surprising amount of variability in total energy consumption. When we attempted to isolate this variability, we discovered that much of the variability was due to the fact that the examples in the batch the model was seeing had different levels of complexity.
     The variability in measurements increased further when we compared across batch sizes as now there was an additional source of variation since larger batch sizes have a larger memory footprint but also can be compressed and transferred more efficiently than smaller batch sizes.
     \end{scode}

\subsection{Component Specific Challenges}

\subsubsection{Accelerators}
When considering accelerators in a system, whether they are on-chip or off-chip can impact energy measurement strategies. On-chip accelerators, integrated onto the CPU mesh, typically do not have a distinct method available to get power measurement that is separate from the total package power/energy consumption. To estimate energy consumption accurately, an appropriate performance to power ratio derived from running a power benchmark on a similar processor should be chosen (see Section \ref{sec:proxies}). In situations where a power benchmark can not be achieved, the socket level power estimation is likely to include the additional power consumed by the on-chip accelerators. For off-chip accelerators, they should be treated as distinct components and estimating their power consumption requires using relevant tools or proxy measurements tailored to the specific accelerator architecture and workload characteristics. 

\subsubsection{Memory}
\label{memory}
Memory power usage may or may not be negligible. it is essential to consider the specific context, workload characteristics, and system requirements when assessing the impact of memory power consumption on overall system energy efficiency and performance.

Memory modules, such as RAM, generally consume relatively low power compared to other components in a computer system, such as the CPU or GPU (though this may depend on the system configuration \citep{ashrae}). Low power consumption is because memory modules primarily consist of static or dynamic semiconductor elements that do not require significant power to maintain stored data. This is especially true during typical operation when the CPU and GPU are actively performing computations and accessing data, while memory is primarily responsible for storing and retrieving data as needed.

\begin{sjob}
\textbf{Real World Learning (S2):}
Cache-Aware Deep Learning Techniques:
Dividing large models into smaller, cache-sized units boost energy efficiency. Ensembles of smaller models often outperform a single large model with the same amount of trainable parameters \citep{ButterEnergyPaper}.
\end{sjob}

However, if the context of our measurement changes from a snapshot of the system at a particular time to the general lifetime of the system, the impact of memory management may not be negligible. While the energy usage of individual memory modules may be low, the aggregate energy consumption of memory subsystems (especially in large-scale server or data center environments with multiple memory modules) can become significant. It may warrant consideration in overall system power management strategies. In some situations, the research question might in fact be an energy comparison of two or more memory systems. Memory energy usage can also vary depending on factors such as memory access patterns, data retrieval frequency, and memory module type (e.g., DDR3, DDR4). High-frequency memory access or memory-intensive workloads may lead to increased power consumption. And even in idle state, memory modules may still consume power to maintain data integrity and readiness for rapid access.

\subsubsection{Networking}
\label{Networking}
The energy associated with networking equipment can be among the most difficult to find. In many cases, networking equipment is not \href{https://ripe87.ripe.net/archives/video/1143/}{power proportional}, meaning that its use of power does not directly or immediately fluctuate with additional work (in the long term, however, increased overall demand does lead to the buildout of more energy-consuming equipment). In scenarios that are not networking intensive, or where you have no options to use an alterntive networking capability, it can be reasonable to leave networking out of scope for measuring. However, batch sizing for the networking capacity you have does make a major difference to the overall component energy consumed.  

\begin{scode}
\textbf{Real World Learning (S3):}
In the case of neural network training, optimizing the data transfer between RAM, CPU and GPU showed significant energy gains. For example, finding the ideal batch size noticeably minimized overall power consumption. If the batch size was decreased or increased from this optimal point, more power was consumed; in the case of small batch sizes, due to more data transfers or in case of large batch size, the additional time to load the data into memory. 
\end{scode}

\section{Proxy Estimations}
\label{sec:proxies}
In situations where a device does not report power data, access to the device’s power registers is restricted or otherwise unavailable, and adding a power meter is not feasible, a proxy estimate must be used. Proxies usually involve estimating consumption based on resource utilization metrics, which, while related to power, is not a power measurement itself. Proxy estimations are different from both direct power measurement and other ways of estimating, because they require relying on \textit{only} indirect measurements of power, typically taken outside the immediate context of measurement. The main difference between other forms of estimates and proxies is that other estimates are based on some aspect of power consumption scaled appropriately to the scope of estimation, whereas proxy measurements rely on indirect energy/power measurements in order to calculate \emph{probable} power consumption. 

For example, power consumption can be estimated by mapping resource utilization percentages to a pre-generated performance to power ratio of the system or a similar one. Resource utilization affects power efficiency in complex ways: a server may draw significant power even when not fully utilized, and tasks requiring 75\% CPU utilization may consume nearly as much power as those requiring 100\% utilization.

\begin{figure}[h!]
    \centering
    \includegraphics[width=0.5\linewidth]{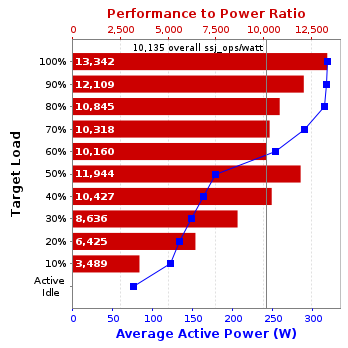}
    \caption{Example of a power to performance ratio for a specific study \citep{standard_performance_evaluation_corporation_specpower_ssj2008_2024}.
    }
    \label{fig:power_loadline}
\end{figure}

A performance to power ratio illustrates the relationship between CPU resource utilization and power draw. This ratio is typically generated using power benchmarking suites such as SPECPower® \citep{lange2009identifying} or the Server Efficiency Rating Tool (SERT®) \autocite{von2018measuring}. An example of such a ratio produced by SPECPower can be seen in \autoref{fig:power_loadline}. Many Original Equipment Manufacturers (OEMs) provide these for servers and other systems to assist with estimations. By monitoring CPU utilization over time and refering to the ratio, energy consumption can be estimated. This is a common industry practice that organizations like Teads and open source projects like Boavizta and CodeCarbon utilize. However, this method introduces a higher margin of error compared to direct measurements because the benchmark used to create the ratio may not accurately reflect the specific workload of the application or the system configuration.

\begin{tipbox}
\textbf{Tip!} If there are multiple performance to power ratios available for your system, choose the one that was bench-marked on a workload most similar to your workload.
\end{tipbox}

When identifying the ratio that is most similar to your particular device, key factors like CPU architecture, CPU clock speed, power efficiency optimizations, thermal design power (TDP), and workload-specific performance are critical. While finding an exact match to your scenario -- especially workload specifics -- may be challenging, selecting a curve with the same architecture and TDP should be a solid starting point for estimations.

For example, let’s consider an Intel (R) Core (TM) i7-7660U CPU with a clock speed of 2.50GHz, used primarily for I/O operations. Start by looking for performance to power ratios provided by the manufacturer or third-party sources (e.g., Tom's Hardware, SPEC, or OEMs like Dell). However, keep in mind that you may not always know the specific workloads used to generate these. As a last resort, you can estimate a performance to power ratio based on TDP or benchmark your own workload to generate a more accurate measurement.

\subsection{Benchmarked Performance to Power Ratios}
It is essential to understand something of how benchmarks are created in order to interpret their outputs. The benchmarking process begins with a calibration phase to determine the system's maximum throughput ($M$). This is achieved by executing workloads designed to engage nearly 100\% of the available compute resources, although this figure may not represent a literal 100\%.

Subsequent to this calibration, benchmarks are run at lower utilization levels (e.g., 90\%, 80\%) by adjusting the throughput to a percentage of the maximum throughput ($M$). For instance, to measure power consumption at a 90\% load, the benchmark is executed with a throughput of $M \times 0.9$. This procedure establishes a relationship between resource utilization and power draw, which can be used to estimate the power consumption of a specific application.

During the execution of the application/job/code you are attempting to estimate the energy consumption of, a background process can be implemented to monitor resource utilization. By applying the ratio derived from the benchmarking, one can estimate the power consumption throughout the application's runtime.

When using data from benchmarks, consider the assumptions used in creating them, which in turn may affect how well they correspond to your own estimation. For example:


\begin{itemize}
    \item The benchmark is run on a bare metal system in a lab setting.
    \item The benchmark is unique to the system/server configuration.
    \item There are few benchmarks for power measurement of AI workloads that generate a performance to power ratio as of this time. The underlying workload of SPECPower® and SERT® is server-side java (i.e. http/https transactional requests). 
    \item Proxy measurements do not comprehensively consider all factors affecting power,  e.g. temperature and the `clock-state' of the CPU.
    \item It can be difficult to isolate per application use and it is possible to miss some aspects of an application that one does not have direct access to. 
    \item If using a virtual machine or docker container it is highly challenging to measure accurately due to the abstraction layers and shared resources involved.
\end{itemize}

Because these ratios depend heavily on hardware configurations (i.e. which peripherals are connected like monitors, what is the processor state, etc.), benchmarks require precise setups that may be impractical to replicate in most experimental or production environments. As a result, they are not easily comparable across systems. Relative power analysis is more feasible. Given this, publicly available performance to power ratios offer similar insights without the need for generating individualized ratios for your system. 

\subsection{Thermal Design Power}
The published Thermal Design Power (TDP) of hardware can also serve as a proxy for power consumption. TDP represents the maximum amount of heat generated by the hardware component, typically measured in watts. It therefore represents "worst case scenario" conditions, where hardware is being run at its upper limit. Because TDP is a thermal specification rather than a direct measure of electrical power, it might not  accurately reflect real-world power consumption, which is affected by workload intensity, operating conditions, and system configuration. Be mindful that the definition of TDP depends on the manufacturer and might not be comparable across the board. Many online tools like \href{https://engineering.teads.com/sustainability/carbon-footprint-estimator-for-aws-instances/}{Teads} (\href{https://medium.com/teads-engineering/building-an-aws-ec2-carbon-emissions-dataset-3f0fd76c98ac}{and others}) use the TDP value to derive the performance to power ratios of CPUs.

 If reliable  proxies are not available for certain components, researchers may need to adjust their approach to estimation. This could involve redefining the research question to ensure comprehensive and accurate estimation of power consumption of all relevant components.
\section{Measurement Interpretation and Analysis}
\label{sec:interpretation}
After measurements or estimates are taken, there is always an interpretation and analysis process which will correlate those measurements to draw data-driven conclusions related to the questions of interest. 
While the reader will have a much greater understanding of the interpretation and analysis required for their specific research question, there are some known  'land mines' we outline below. 
    
\subsection{Mapping metrics to objectives} 

Prioritizing known metrics and measurement techniques when possible can improve reliability and interpretability, but in practice the research questions may influence the availability of said metrics. Chosen metrics may be a function of availability rather than accuracy.



Ensuring the scoping remains consistent helps to ensure  that comparisons can be made and trusted. It is often useful to define the scope to include how the different hardware components are utilized and interact. For example, is data processing, I/O operations, networking, etc. in the scope of measurement? If so, how does energy consumption and resource utilization change during different phases of the workflow? Does there need to be differentiation between system level and component level power consumption? Also consider whether you need to measure on multiple different hardware configurations for a more holistic understanding of how the software performs in various environments.

Similarly,  clarify whether you are interested in overall energy consumption or specific components contributing to it, like CPU usage, memory usage, disk I/O, and network activity. Define a unit of work that will form the basis of your test, like energy consumption per inference or per training epoch. Then choose a metric, which could include total energy consumption, average power usage, or energy efficiency metrics like performance per watt, ensuring that they are consistent and comparable across different hardware configurations.  

The level of your analysis (whether system, job, or code level)  informs the kinds of conclusions that can be drawn. Some examples are below.\\

\vspace{0.7em}

\textbf{Scenario 1: System level analysis}
Scenario 1 involves the comparison of system level choices that may affect multiple jobs and their interaction. However, it is much more difficult to disambiguate the impact on specific parts of a job. Typical use cases include investigating the impact of schedulers, default system settings, job policies, or cooling hardware. 

\textbf{Scenario 2: Job level analysis}
Typically when comparing job energy or power, we are looking to see how differences in the execution environment (algorithm, data, software, operating system, hardware, configuration, etc.) change the energy or power used.
For example, one job might train a wide and shallow neural network while another trains a deep and narrow one on the same dataset; how much energy does each one require to reach a given accuracy?
Or, we might compare the same workload on two different hardware configurations to understand which is more efficient, and if there is a time and energy trade-off.
Such investigations can be used to optimize the execution environment-- choosing the best algorithm, software, hardware or configuration for a workload.

\begin{sjob}
Real world learning (S2): Accurate job level analysis relies on our ability to control for sources of noise and bias in our measurements. In this case we used repetitions of identical workloads on the same and different hardware configurations to characterize the nature of sample noise and the differences in runtime and energy consumption between hardware configurations.
\end{sjob}

\textbf{Scenario 3 : Code level analysis}\\
Comparisons that enable understanding where efficiency can be gained, whether for hardware or software often require code level power/energy analysis. Similar to a profiler for faster code execution, this type of scenario is useful to improve the efficiency of \emph{specific} aspects of the overall job and thus cannot readily be used to support job level conclusions.
This often means metrics and objectives should be constrained to conclusions related to immediate execution. S1 power measurements cannot be used to reliably track power usage over time (e.g. this approach would be unreliable for estimating the energy consumption of inference vs training of a deployed neural network) nor can it be used to show improvement in efficiency over time as showing an immediate efficiency in one part of the code does not mean the overall job level is more efficient.

Especially in the case where you are interested in comparing specific aspects of a larger job to each other or optimizing the efficiency of one aspect of the code, consider scoping the power measurements to natural workflow phases (e.g., data processing, computation, I/O) or algorithmic units (e.g., choose the epoch level rather than specifically focusing on the back-propogation step within the epoch). 
You might examine resource utilization across hardware components (CPU, GPU, memory) during task execution, and differentiate between system-level power consumption and component-level power consumption.

\begin{scode}
Real world learning (S3): While we were able to show with high reliability that one epoch of a transformer was more efficient than a single epoch of an LSTM, we could not generalize this to whether training a transformer model took overall less power than the LSTM model as the architectures have different learning trajectories and thus require different number of epochs to train. Only when assessing the overall job level could we compare the architectures directly. Even with this limitation, we were able to say that transformers took less time than LSTMs in pre-processing the batches, indicating, at least for our small example, an opportunity for improving LSTM efficiency by focusing on the data pre-processing.
\end{scode}
    
\subsection{Assessing measurement influences} 
The factors that influence measurement depend on where the computation is  being run. Bare metal systems provide the most reliable estimates of power and the most user control. On the other extreme, measuring energy consumption on virtual machines is notoriously difficult and challenging, yet this setting is very common, and therefore nevertheless important.
Your specific operational environment will dictate what could influence the measurements, such as background tasks, shared resources or resource limitations, and environmental conditions. 
In AI models this could also be hardware platforms used, model architectures selected, or algorithmic efficiency. 

More detailed and narrow comparisons (e.g. when looking to improve efficiency of a particular part of the code (S3)) are more prone to influence from background processes and setup conditions (i.e. which dataset is used, compiler settings, python package versions etc). Further,
very short pieces of code (lasting <0.1 seconds) are strongly influenced by the power draw of measuring tools themselves.
    

\subsection{Baseline establishment} Comparisons typically require a baseline of some kind.  This could be power consumption under typical workloads, un-optimized algorithms, or industry benchmarks. Only compare results that were measured with the same methodology and scope. If in one measurement you included the cost of networking I/O, do not compare it with the measurements excluding it. 

This might require creating a standardized testing environment to ensure fair comparison, as much as is feasible. This environment should include the same operating system and workload conditions for both systems. If you can't replicate the exact hardware configuration, try to match it as closely as possible in terms of performance characteristics and isolate conclusions to efficiency of the software rather than hardware. 
Even with the same setup and workload, there will still be noticeable variation, thus repeating the same measurements multiple times will lead to more reliable power/energy estimations. 


\subsection{Reporting}
Whether in a research publication or as part of a model card, choosing what and how much to report can also be a challenge. While maximal transparency is a good goal, what that looks like practically can vary, depending on the needs of the stakeholders to whom you are offering transparency. Below are some key elements of transparent reporting ( considerations about emissions reporting is in  Section \ref{sec:Energy_and_Emissions}).  

\textbf{Hardware Characteristics:} Start by clearly outlining the hardware components involved in the analysis. Specify the types of components measured, such as CPU, GPU, FPGA, accelerators, memory, and storage. Provide detailed information about the make, model, and key hardware specifications, including clock speed, number of cores, and memory size. Additionally, describe how the components are configured within the system, such as server setup, cooling solutions, and overclocking.

\textbf{Software Characteristics:} Detail the software environment and configurations used during the power analysis. Provide context to the measurements by considering the specific aspect of the application under evaluation. For instance, when the application focuses on inference tasks, consider reporting energy consumption measurements relative to the number of floating-point operations executed per second. Explain any optimizations made to the code to enhance performance and efficiency, such as algorithmic improvements and memory management techniques. Describe if, and how, parallel processing techniques are utilized, like multi-threading or distributed computing. Include information about the operating system, drivers, libraries, and any other relevant software.

\textbf{Measurement Methodology:} Describe the methods used to measure power consumption, listing the tools and devices used for measurements,. Explain the setup and conditions under which measurements were taken, including idle versus load conditions and isolation of specific devices. Discuss any calibration steps or accuracy considerations for the measurement devices.

\textbf{Additional Considerations:} Provide contextual information to help interpret the power/energy measurements. Clearly state the units of measurement, like Watts or Joules. Include details such as average and peak power consumption, comparisons with benchmarks, and idle power usage to help understand the measurements in various operational contexts when applicable.

\textbf{Sources of Error:} Acknowledge potential sources of error and limitations in the measurement process. Discuss any assumptions made during the measurement process and, the potential limitations of the methodology, which could include environmental conditions, measurement tool precision, or system-specific factors.

    

\chapter{Energy \& Emissions}
\label{sec:Energy_and_Emissions}
A common objective in tracking power and energy consumption is to correlate these metrics with greenhouse gas emissions. However, this process is complex. Converting energy consumption data into carbon emissions estimates, requires accounting for the carbon intensity of the energy sources used. Carbon intensity is defined as the amount of carbon dioxide (CO$_2$) emissions produced per unit of energy generated and varies depending on the energy source, such as coal, natural gas, or renewable energy. Power grids report their carbon intensity, and third-party organizations, such as WattTime and ElectricityMaps, provide tools to access regional carbon intensity data. Forward-looking estimates under different scenarios can be obtained via tools like Cambium \citep{gagnon_cambium_2023}.


The timing and location of energy use significantly influences carbon emissions estimates. Power grids typically employ a mix of energy sources, with the proportion of renewables versus fossil fuels fluctuating based on demand and availability. During peak demand periods, grids may rely more heavily on fossil fuels, resulting in higher carbon intensity (though not always).


It is increasingly common to include disclosures regarding energy use and greenhouse gas emissions in model cards and scientific publications. While this represents a positive and significant development, it is also important to correctly interpret common practices, and provide interpretable information. Specifically:

\begin{enumerate}
    \item Going from energy used to emissions involves having a clear understanding of the time and place where energy is used. Often model developers report emissions based on the last training experiment using the average for their region, which is helpful for comparing models. However, changing where and when the work is being done can have a major impact on the actual emissions, because the carbon intensity of the power grid where the work is done changes. Therefore, the actual emissions created in the process of development, experimentation, and productization of the model might be quite different. A disclosure does not necessarily predict the emissions of the model in future situations. 
    \item Transparency efforts should be attuned to the expectations and needs of stakeholders. Someone using a model in the cloud via an API is going to care about the energy associated with inference, which in turn depends on the location of that model in the cloud. On the other hand, if someone is using an off-the-shelf training recipe to train a model with new data, information about the energy used in previous training runs can be useful as a rough guide to what they might expect. The carbon footprint, on the other hand, might not serve as a guide, because subsequent trainings depend on the users' location, not the model providers'.   Measures for action—(is this software going to cost me more energy over that?) and measures for accounting, which typically strive for fuller life cycle analyses, are two very different matters.   
   
    \item Prioritize the metrics you wish to disclose according to how reliable they are for someone using your model outside of your context. For example, in a situation when you are releasing a model into open source for other developers to use, you know nothing about other developers’ environments. Therefore, the most reliable metric could be runtime over hardware  (e.g., X hours on Y amount of these resources). Even though this is a proxy for energy and not a direct measure, it is the clearest, most interpretable signal. If the original developer had an inefficient software stack, runtime by hardware would constitute the worst-case scenario for the receiver. If the original developer had a very efficient software stack, any worse power performance would be the responsibility of the receiver. From there, it is possible to also include an estimate of energy, but do not report energy without reporting runtime over hardware.
\begin{tipbox}
\textbf{Tip!} Because measurement methods are not yet standarized, in a model card, runtime on hardware (e.g., X hours on Y resources) can be a more interpretable signal to other developers about potential energy use than an energy measurement. 
\end{tipbox}
    For similar reasons, in that same scenario do not report emissions without reporting runtime over hardware used, power estimates, software location (which affects emissions), and how those emissions were derived e.g from a yearly average or based on real-time actuals and gird location. In a model card, yearly averages for the location where the compute was run are more useful for comparability between models, even if less accurate from an embodied emissions perspective. On the other hand, in a situation where the user is calling an API to generate an inference in the cloud, emissions generated by non-averaged actuals gathered in real time from a service like WattTime or Electricity Maps, are more accurate.  Similarly, real-time emissions data becomes more appropriate when tracking embodied emissions over the course of development and/or use. 
    
    \item Consider whether you want to factor in the carbon cost of procuring the hardware used, or other factors like water usage effectiveness (WUE) in your reporting. Most carbon estimation tools exclude these, and only focus on direct energy use, but they do paint a fuller picture. Life Cycle Assessments (LCAs), sometimes provided by manufacturers, reveal the carbon cost of producing hardware from manufacturing to end-of-life. How you map hardware lifecycle emissions onto a specific use of that hardware requires skills beyond the scope of this paper, such as decisions about attributional versus consequential greenhouse gas accounting (see \cite{su13137386}).   
\end{enumerate}
\chapter{Discussion/Conclusion}

As shown, accurate measurement and estimation of energy present significant challenges, from noise and inaccuracy issues to scoping and commensurability challenges. Current methodologies often result in misleading comparisons because they lack comprehensive information and standardized practices. That does not mean, however, that it is impossible or too difficult to be worthwhile. In a climate crisis, useful, even if imperfect measurements, are vital. In this paper, we have outlined ways that measurements can mislead, to help others gain greater familiarity with this topic. However, the expertise of individual researchers is not enough.  Given the exponential rise in energy consumption and carbon emissions driven by increasing computational demands, it is crucial to enhance our understanding and optimization of computational energy use more broadly. 

Rising to that challenge entails three things. First,  enhancing the toolchain for computational energy measurement is essential for bridging gaps in our current understanding and achieving reliable, cost-effective estimates. We encourage the research community to advance the accuracy and consistency of energy measurements and estimates, particularly when direct measurements are impractical, particularly at system and component levels. It is vital that advances ensure that that these methods support clear and detailed comparisons of algorithmic and hardware choices.

Second, reporting of energy and emissions costs should become part of researchers' and developers' standard workflow and expectations.  That in turn requires much better standards than we have today. While there are non-standardized emissions disclosure statements emerging within ML research communities, the absence of standardized measurement protocols represents a critical challenge that impedes the ability to compare results across different systems and studies. Inconsistencies across platforms and computing environments undermine the reliability and validity of power consumption estimates, highlighting a substantial gap in both research and practice. Researchers and engineers are urged to develop and advocate for standardized methods that enable accurate and generalizable measurement of computational energy consumption across various operating environments. Such methods should facilitate robust comparisons between diverse hardware and software.

Third, the disconnect between the perspectives of data center owners and software developers and researchers needs to be overcome. In this paper, we have primarily taken the perspective of developers writing code, who often rely on those who manage a data center to access energy data, and can experience limitations. Data center managers, on the other hand, are more concerned with developers' collective patterns of behavior and its impact on resource use. Developers' primary concerns, like making software more efficient, involves a set of actions that are almost orthogonal from the actions that would enable the data center to make more effective use of energy (for example, maximizing use of a single processor before involving the next). There is an urgent need for interventions, whether technical or organizational, that facilitate better shared responsibility for energy stewardship.   

The increasing focus on energy efficiency, driven by advancements in artificial intelligence and rising computational demands, underscores the urgency of addressing these measurement challenges. Progress in technology and AI relies on overcoming these barriers to support sustainable development. Continued research and development in this domain are vital for obtaining precise and actionable insights into power consumption. That research is twofold. Energy specialists should continue improving the state of the art and enhancing the toolchains that deliver reliable, cost-effective, and usable estimates. In parallel, ML researchers whose core research is not energy might find that incorporating an energy lens as part of their research could open up new research pathways and steer ML subfields towards better energy use.  

\appendix

\chapter{Appendix 1: FAQ}
\label{faq}
\begin{itemize}
    \item \textbf{Is there a benchmark for power measurement?}
    
  There isn't a universal benchmark for power measurement applicable to all situations, but benchmarks are emerging all the time (see for example \cite{poess2010energy}). Tools like SERT® and SPECPower® are in common use, including use by ENERGY STAR®  certifications, but there the focus is on hardware measurement. Power benchmarks for ML workloads are being developed by \href{https://mlcommons.org/working-groups/benchmarks/power/}{ML Commons.} 

    \item \textbf{Why is there no single standard for how to measure power? It’s just physics at the end of the day.}

    Physics is universal but system setups and operational conditions are not. Diverse hardware configurations, software workloads, and environmental factors affect power consumption in ways that cannot always be perfectly controlled. Appropriate scoping also varies with the purpose, which makes one standard scope or measurement method difficult to align on. The \href{https://en.wikipedia.org/wiki/Coastline_paradox}{Coastline Paradox} also complicates the problem. This is the mathematical principle that says that the more detail with which someone attempts to measure a real-world phenomenon, like a coastline, the bigger the tally is. It does not bottom out into a settled, final answer.  With that said, there still is an urgent need to create standardized measurement protocols that support clear and detailed comparisons of algorithmic and hardware choices. 

    \item \textbf{How do I go from energy estimates to carbon estimates?}

    To convert energy estimates into carbon estimates, you need to consider the carbon intensity of the energy sources used to generate that energy, see Section \ref{sec:Energy_and_Emissions} for more details. Note that when using real-time carbon intensity data, the carbon intensity values reported could have a different sampling rate than your power measurements. For instance, you could be measuring your power every 1 second but the carbon intensity values you receive may only be every 5 minutes. Depending on your situation, you can either upsample the carbon intensity values or downsample the power measurements.

    \item \textbf{Can the published TDP (Thermal Design Point) of hardware help me estimate power?}

    Yes, the published Thermal Design Power (TDP) of hardware can provide a rough estimate of power consumption, as long as some caveats are taken into account. See Section \ref{sec:proxies} for more details.
    
    \item \textbf{How does utilization affect power consumption?}

    Utilization, or the level of activity or workload processed by the hardware, directly affects power consumption, however, this relationship is usually nonlinear, see Section \ref{sec:proxies}.
    
    \item \textbf{When do you really need a ‘clean,’ out-of-the-box, bare metal system?}
    
    For benchmarking, performance testing, or evaluating hardware, measuring power on a bare metal system is preferable. However, if you are aware of the final deployment environment for your application and it won't be running on a clean system, mimicking the environment as closely as possible is necessary to understand your application's power consumption accurately.

    \item \textbf{When should you measure at the wall versus specific devices?}

    Measuring power consumption at the wall, also known as whole-system or out-of-band power measurement (see Section 2.4), is particularly useful in several scenarios:
    \begin{itemize}
        \item \textit{Consumption of the entire system:} If the goal is to create a comprehensive view of energy usage of a system, it is sufficient to measure at the wall as it will include all components (e.g., CPU, GPU, memory, storage, peripherals.
        \item \textit{System Optimization:} For optimizing energy efficiency at the system level, measuring power at the wall offers insights into the combined impact of hardware configurations, software workloads, and system settings on energy usage. This information is valuable for identifying opportunities to reduce power consumption, improve performance-per-watt metrics, and optimize resource allocation.
        \item \textit{Environmental Conditions:} Measuring power at the wall accounts for inefficiencies in power conversion, distribution losses, and environmental factors (e.g., temperature, humidity) that can affect overall energy consumption. 
        \item \textit{Cost estimation:} If energy expenses are a significant concern, measuring power at the wall enables accurate estimation of energy costs based on local utility rates and facilitates informed decisions regarding resource allocation and efficiency improvements.
        \item \textit{Regulatory compliance:} In regulated industries or regions where energy efficiency standards or carbon emissions regulations apply, measuring at the wall can help organizations obtain accurate measurements of total energy usage.
    \end{itemize}

    \item \textbf{When other applications are using the same resource, is there a ‘hack’ for estimating power of your application?}

    Unfortunately, there is no easy hack for estimating power of your application when sharing resources. However, there are a few methods we can employ.
    \begin{itemize}
        \item \textit{Proxy metrics:} If direct  measurement isn't feasible, consider using proxy metrics that correlate with energy consumption. For example, CPU utilization, memory usage, or disk activity can serve as proxies (see Section \ref{sec:proxies}).
        \item \textit{Simulation:} Develop models or simulations to estimate energybased on known application characteristics and system parameters. By modeling the power-performance relationship of your application and its interaction with shared resources, you can predict energy consumption under different scenarios and workload conditions.
        \item \textit{Isolation:} As much as possible, isolate your application from competing workloads or shared resources to minimize interference and simplify energy estimation. By controlling the execution environment and limiting external factors that could impact energy consumption, you can obtain more accurate estimates of your application's power requirements.
        \item \textit{Collaboration:} Collaborate with system administrators or other users sharing the same resource to gather insights into resource utilization patterns and energy consumption trends.
    \end{itemize}

    \item \textbf{How does the length of the process affect what is being measured?}

    Real-world systems are dynamic and subject to variability in workload, environmental conditions, and system state over time. Longer processes may be exposed to a wider range of system conditions and environmental factors, leading to increased variability in performance and power consumption measurements. They may also encounter transient effects or anomalies during execution, such as sudden spikes in resource usage, intermittent bottlenecks, or temporary system failures. These transient effects can distort performance and power measurements, particularly if they occur sporadically or unpredictably during the process's runtime.

    \item \textbf{I am using a virtual environment. Should the energy it takes to run the virtualization software count?}

    The answer depends on the task you are trying to measure. Think about how energy usage is allocated in your context. Virtualization software typically consumes resources such as CPU, memory, and disk I/O. If you allocate the energy usage of these resources to the virtualized workloads, it may not be necessary to separately account for the energy used by the virtualization software.

    If the primary focus of your energy measurement is to evaluate the energy efficiency of the workload or tasks executed within the virtual environment, it may be more appropriate to exclude the energy consumption of the virtualization software. However, if you aim to conduct a comprehensive analysis of the overall energy usage of the entire virtualization infrastructure, including the energy consumed by the virtualization software becomes relevant. The broader perspective allows you to assess the efficiency of the entire virtualization stack, including hypervisors, management software, and associated infrastructure components.
    
    Either way, energy measurement in a virtualized environment can be less reliable. For example, if 
 an application is running inside a guest OS, the guest OS could be moving across different cores, sockets or even nodes. It remains important to clearly document which components you are including in your measurement and ensure consistency in how you normalize the measurements.

    \item \textbf{My hardware uses hyperthreading and I am getting a utilization number that is higher than 100\%. What is happening? }

    Exceeding 100\% utilization in systems utilizing hyperthreading is a common occurrence due to the way hyperthreading technology works. Hyperthreading, also known as simultaneous multithreading (SMT), enables a single physical processor core to execute multiple threads simultaneously, presenting itself as multiple logical cores to the operating system. When a processor with hyperthreading is fully utilized, each logical core may be executing instructions concurrently, resulting in a utilization percentage exceeding 100\%.

    To address this issue and estimate power consumption accurately, consider the following steps:
    \begin{enumerate}
        \item \textit{Understand hyperthreading:} Hyperthreading enables better utilization of CPU resources by allowing multiple threads to execute on a single physical core concurrently. As a result, the operating system may report utilization percentages greater than 100\% when all logical cores are fully utilized.
        \item \textit{Calculate true utilization:} To estimate energy consumption accurately, it is essential to calculate the true utilization of the physical cores rather than relying solely on reported utilization percentages. Divide the reported utilization by the number of physical cores to obtain a normalized utilization value per physical core.
        \item \textit{Adjust power estimation:} Once you have obtained the normalized utilization per physical core, use this value to estimate power consumption more accurately. Power estimation models or power profiling tools may need to be adjusted to account for hyperthreading and ensure that power estimates reflect the actual workload on the physical cores.
        \item \textit{Consider hyperthreading overhead:} While hyperthreading can improve overall system throughput, it also introduces overhead due to context switching and resource sharing between logical cores. Consider the additional power consumption associated with hyperthreading overhead when estimating power for systems utilizing hyperthreading technology.
        \item \textit{Validate with HW measurements when possible:} Validate power estimates obtained through software-based methods with direct hardware measurements using power meters or monitoring tools. Hardware measurements provide the most accurate assessment of power consumption and can help verify the reliability of software-based estimation techniques.
    \end{enumerate} 

    \item \textbf{Why bother with measuring at all if the data center I am using consumes so much more energy than my one application?}

    There are many layers between the resources that software directly uses, and the infrastructure on which it relies. In most cases, it is unrealistic to expect that the amount of energy saved in reducing a workload is the same thing as avoided energy as it comes off of the grid and into the building. 

    However, in the long term, the opposite is also true: large software loads increase demand for energy, as evidenced by the proliferation of press reports about energy providers' limited ability to  accommodate the AI boom, in some cases resorting to fossil fuels to meet demand. In the long term, one key reason to keep track of energy is to avoid unnecessary power use becoming a kind of technical debt as code is passed from one person to the next, influencing the aggregate. Another is to ground developer intuitions about what is efficient code based on data and not guesses. What we think might constitute an optimization of code might not actually be an optimization from an energy perspective.    
    
    
    \item \textbf{What is Power Usage Effectiveness (PUE)?}
    
    PUE is a metric commonly used in data center operations to evaluate the energy efficiency of a facility. PUE is calculated by dividing the total amount of energy consumed by the data center facility by the energy consumed by the IT equipment within that facility. A lower PUE value indicates higher energy efficiency, as it implies that a smaller proportion of energy is being consumed by non-IT equipment and support infrastructure relative to the energy consumed by the IT equipment itself.

    \item \textbf{I am measuring at the application or code level. Should I take the energy costs of networking into account?}
    Only in scenarios where software is known to be particularly networking-intensive, see Section \ref{Networking}.
\end{itemize}

\chapter{Appendix 2: Scenario Walkthroughs}

Throughout this paper, we referred to three scenarios outlined in the General Approach section (i.e., system, job, and code level  scenarios). Here is a brief account of each of them, for those desiring a more complete set of examples in one place.

\section{Scenario 1 (System Level): Evaluating the impact of HPC system scheduler on whole system power variability}
NREL researchers were interested in understanding the impact of more effective job scheduling on reducing total system power variability, specifically to investigate the potential impact of peak shaving \cite{bugbee2017prediction}. To investigate, a dataset was collected of power usage for every job scheduled over the course of several years. From this dataset researchers were able to make statistical descriptions of workload patterns and system power use. These workloads were then used to estimate the impact of the scheduling algorithm, using the existing dataset as a baseline 

\textbf{Hardware/Software Characteristics:} The principal HPC machine analyzed in this work is Peregrine, a Hewlett-Packard (HP) system composed of 1440 Intel Xeon nodes incorporating SandyBridge and IvyBridge architectures. Two hundred and eighty-eight of these nodes are accelerated by Xeon MIC Phi co-processors.

\textbf{Measurement Methodology:} The compute nodes are instrumented with HP’s integrated Lights-out (iLO) out-of-band system, with node-level power and thermal data routed through an external server. Aggregated mean power measurements for jobs active in non-overlapping 15-min windows was compiled for one month (April 2015). 

\textbf{Additional Considerations:} As with any data from a large complex system, there is an unavoidable portion of noisy data which must be identified and isolated in analysis. Among the 1440 iLO chips, some fraction habitually record zero values or out of range estimates. Any time series with zero variance (constant readings) are excluded a priori. Despite being important from a systems perspective, we neglect the impact of shared resources including power use by cooling, storage, and communication. This was due to the difficulty in disentangling these resources for an individual job, and the minimal impact they have on the overall work's conclusion due to relatively low power draw when compared to the compute resources.

\textbf{Error and variation}: While aggregating any individual job to mean power usage in 15 minute windows is an oversimplification, it averages out at a statistical level across the entire system.

\section{Scenario 2 (Job Level): Comparing the energy cost of neural network architectures and hyperparameter selections using an instrumented HPC system}

\textbf{Hardware Characteristics:}
Each experiment was executed using all CPU cores or all GPUs of a single HPC node.
Each node had one of four hardware configurations as described in Tables \ref{tab:s2_hpc_nodes} and \ref{tab:s2_spec_comparison}.
CPU-based experiments ran only on CPU node types, and GPU-based experiments ran only on GPU node types.

\begin{table}[h!]
    \caption{Specifications of the HPC nodes used in Scenario 2. Table adapted from \cite{ButterEnergyPaper}.}
    \label{tab:s2_hpc_nodes}
    \centering
    {
    \setlength\tabcolsep{1.0mm}
    \begin{tabular}{|c|rcrcr|r|}
        \midrule
         Name & \# & CPU & RAM & GPU & SSD & Idle Power\\
         \midrule
         cpu1 & 1800 & 2x Xeon 6154 & $\SI{96}{\gibi\byte}$ & - & $\SI{1}{\tera\byte}$& $\SI{220}{W}$\\
         cpu2 & 720 & 2x Xeon 6154 & $\SI{192}{\gibi\byte}$ & - & $\SI{1}{\tera\byte}$& $\SI{230}{W}$\\
         cpu3 & 38 & 2x Xeon 6154 & $\SI{768}{\gibi\byte}$ & - & $\SI{1.6}{\tera\byte}$& $\SI{309}{W}$\\
         cpu4 & 10 & 2x Xeon 6154 & $\SI{768}{\gibi\byte}$ & - & $\SI{25.6}{\tera\byte}$& $\SI{388}{W}$\\
         gpu1 & 40 & 2x Xeon 6154 & $\SI{768}{\gibi\byte}$ & 2x V100 & $\SI{1.6}{\tera\byte}$& $\SI{374}{W}$\\
         gpu2 & 10 & 2x Xeon 6154 & $\SI{768}{\gibi\byte}$ & 2x V100 & $\SI{25.6}{\tera\byte}$& $\SI{403}{W}$\\
         \midrule
    \end{tabular}
    }
\end{table}
\begin{table}[h!]
    \caption{CPU and GPU specifications for Scenario 2. Table adapted from \cite{ButterEnergyPaper}.}
    \label{tab:s2_spec_comparison}
    \centering
    {
        \renewcommand\footnoterule{}     
        \begin{tabular}{|c|c|c|}
            \hline
            \textbf{Property} & \textbf{Xeon Gold 6154} & \textbf{V100 PCIe} \\
            \hline
            \DeclareSIUnit\core{core}
            \DeclareSIUnit\sm{\textsc{SM}}
            Manufacturer & Intel & NVIDIA \\
            TDP & $\SI{200}{W}$ & $\SI{250}{W}$ \\
            Base Frequency& $\SI{3.0}{GHz}$ & $\SI{1.2}{GHz}$ \\
            Boost Frequency & $\SI{3.7}{GHz}$ & $\SI{1.53}{GHz}$ \\
            Memory Bandwidth & $\SI{128}{\gibi\byte/s}$ & $\SI{897}{\gibi\byte/s}$ \\
            Memory/Package & $\SI{48}{\gibi\byte}$-$\SI{384}{\gibi\byte}$ & $\SI{16}{\gibi\byte}$\\
            Cores/SMs & 18 & 84 \\
            L1 Data Cache & $\SI{32}{\kibi\byte/core}$ & $\leq$ $\SI{96}{\kibi\byte/SM}$\\
            L2 Cache & $\SI{1}{\mebi\byte/core}$ & $\SI{6}{\mebi\byte}$ shared \\
            L3 Cache & $\SI{24.75}{\mebi\byte}$ shared & - \\
            FP32 Vector Width & 16 (AVX-512) & 64 FP32 cores/SM\\
            Register File Size & $\SI{2}{\kibi\byte/core}$ & $\SI{256}{\kibi\byte/SM}$\\ 
            \hline
        \end{tabular}
        \vspace{-2ex}
    }
\end{table}

\textbf{Software Characteristics:} 
41,129 individual experimental runs were executed, covering 30,582 combinations of 13 datasets, 20 sizes (number of trainable parameters), 8 network shapes, and 14 depths.
Each experiment trained the network for 3,000 training epochs, evaluating the test and validation sets after each training epoch.
CPU runs used Python 3.9.10, Intel Tensorflow 2.8.0 using OneDNN.
CPU threads were each pinned to a separate core, and SMT was disabled.
OneDNN's default thread allocation policy was used.
GPU runs used Python 3.10.5, Tensorflow 2.8.1, cuDNN 8.1.1, CUDA 11.2, NVIDIA driver 460.32.03, Linux Kernel 3.10.0 using Tensorflow's parameter mirroring strategy.
For GPU runs, one Tensorflow CPU thread was allocated to each GPU and pinned to a different core of a single CPU socket.

\textbf{Measurement Methodology:}
Energy data was collected using Hewlett-Packard Enterprise Integrated Lights-Out (iLO) chip and were recorded once per minute.
The iLO measurements measure the actual power delivered to the compute node, but do not capture overheads such as cooling or power distribution and transmission losses.
Experiment duration ranged from approximately 30 minutes to 16 hours, and many experiments were repeated multiple times.
The measurement sampling rate allows relatively clear analysis of the total, setup/tear-down overhead, and per-epoch energy costs.
However, more specific relationships below the per-epoch resolution are difficult to ascertain without more detailed measurements including a higher measurement rate and records of the start and stop times of each phase of the experiment (e.g. when an epoch began, when it ended, when training evaluation began and ended, etc).

\textbf{Additional Considerations:}
Measurements were taken in Watts and were instantaneous power measurements.
Total energy (Joules) was estimated by integrating a zero-order interpolation of the power measurements over the duration of experiment execution.
Therefore, it is possible that power variations at higher frequencies than half of the sample rate are not well-captured by this method.
For example, a one-second spike in power consumption could be mistaken for a minute-long power draw.
This source of sample bias was controlled for by running experiments for many repetitions of the same calculations, such that many measurements of the same outer loop were taken, and many experiments were also repeated multiple times.

To control for varying hardware configurations, we used the second percentile of the power draw of each node class over a six-month period to standardize the active-idle power draw differences between CPU and GPU node types.
For example, cpu3 nodes had a 2nd percentile power draw of 220W and cpu3 nodes measured 309W.
To standardize results to the cpu1 hardware, we subtracted the 89W difference from all cpu3 experiment measurements.
These offsets were confirmed to be consistent with the data collected using a least-squares fit of active-idle offsets using all experiment repetitions executed on multiple node types.

\textbf{Sources of Error:}
Our efforts to standardize energy consumption and account for power fluctuations not captured by the measurement rate leave some degree of residual error.
Despite these control efforts, comparing repetitions of the same experiment on multiple nodes of the same configuration, we observed some per-node variations in power measurements.
Due to manufacturing, thermal, and electrical variations each node has a slightly different power consumption profile and, in particular, idle power draw.
And, each iLO device has some calibration error contributing to per-node measurement differences.

\section{Scenario 3 (Code Level): Optimizing part of the code to minimize overall power
required for executing the full code}
We conducted an efficiency comparison between Transformer architectures and Long Short-Term Memory (LSTM) architectures, focusing on their power consumption during data processing and back-propagation. 
We were interested in low-level aspects of each architecture such as the power consumed in data processing and back-propagation as well as at the level of training epochs. Specifically, we were interested in whether Transformers and LSTM architectures had different opportunities for power optimization. We were not directly interested in whether training from end to end was more efficient in either architecture (the clear winner in our experiments was the Transformer architecture) but rather if small changes in the model architecture optimization could impact power consumed during training and if those points of power optimization were the same across each architecture.

\textbf{Hardware and Software Characteristics} Our study evaluated low-level aspects of each architecture, including the potential for power optimization. The software environment included careful monitoring of power consumption of one CPU and one GPU, with measurements sampled at varying frequencies (0.01 to 0.1 seconds). We also ensured consistent sampling rates within each experimental run to accurately convert power estimates into energy consumed.

\textbf{Measurement Methodology}
Initially, capturing the power consumption during CPU to GPU data transfer proved challenging. To address this, we conducted measurements on a dataset that fully resided on the GPU, eliminating the impact of data transfer power costs.

Power consumption measurements were taken using microWatts as the unit, given our focus on capturing small-scale energy usage. We used a power curve generated by SPECPower for the relevant processor, estimating CPU energy consumption based on resource utilization. For GPU energy consumption, we employed a Python wrapper to process Nvidia-smi outputs. However, we encountered challenges in ensuring that CPU and GPU power estimations aligned, as they occasionally reflected different stages of the neural network processing (e.g., forward-pass vs. backward-pass). 

\textbf{Additional Considerations}
We observed substantial error and variation in power consumption across different systems, configurations, and hyperparameters. Notably, factors like model size, the number of layers, and sequence length significantly influenced energy consumption, even when controlling for the number of trainable parameters. Data preprocessing and loading methods, especially those involving multi-threaded dataloaders, also impacted power usage. Moreover, numerical variation within epochs introduced further complexity, as different batch sizes and the compressibility of data batches contributed to fluctuating power estimates. These variations underscored the importance of considering operational contexts when interpreting power measurements.

\textbf{Sources of Error}
Throughout our experiments, we encountered various sources of error and variation. Model architecture and hyperparameters were significant contributors, with unexpected impacts on power consumption. Additionally, the data preprocessing pipeline and dataloaders introduced variability that was challenging to control. We also identified numerical variation within epochs as a source of error, potentially linked to the varying complexity of data batches. These findings suggest that linear models may be insufficient to capture the complex interactions between architecture decisions, hyperparameters, and power consumption, prompting further investigation into these dynamics.

\chapter{Appendix 3: Tools}
Below are some common tools that can help measure power. Additional, the Open Compute Project \href{https://www.opencompute.org/w/index.php?title=OCP_Sustainability_Project}{(OCP}) facilitates an ecosystem of relevant workstreams both in instrumentation and in other areas of sustainability beyond power.

\begin{enumerate}
    \item Direct power measurement:
    \begin{itemize}
        \item \textbf{EMON:} Energy Monitoring (\href{https://www.intel.com/content/www/us/en/content-details/686077/emon-user-s-guide.html}{EMON}) is a low-level command-line tool leverages counters from hardware Performance Monitoring Units (PMUs) to collect performance monitoring events to profile application and system performance. 
        \item \textbf{RAPL:} Running Average Power Limit (RAPL) is an Intel technology for monitoring and managing power consumption in real-time across various hardware components, including the CPU, DRAM, and GPU.
    \end{itemize}

    \item Indirect power measurement:
    \begin{itemize}
        \item \textbf{top:} A command-line tool used in Unix-like operating systems to monitor system processes and resource usage in real-time. It provides a dynamic display of CPU, memory, and other system metrics, allowing users to identify processes consuming the most resources and manage system performance accordingly.
        \item \textbf{psutil:} A Python library that provides an interface for retrieving information on system resources and processes. It allows users to access CPU, memory, disk, network usage, and other system information programmatically, making it useful for system monitoring and management tasks in Python scripts and applications.
        \item \textbf{PowerTOP:} A software tool to measure, diagnose, and minimize a computer's electrical power consumption by enabling power saving modes in the userspace, kernel, and hardware.
    \end{itemize}

    \item Software-based energy reporting tools:
    \begin{itemize}
        \item \textbf{PyRAPL:} A python wrapper around RAPL.
        \item \textbf{Powerstat:} A wrapper around the RAPL interface. It provides an average power usage metric, simplifying the interpretation of power data.
        \item \textbf{Perf}: Measures power at different granularities, including cores, RAM, and packages. Despite lacking support for NVIDIA GPUs, Perf is a versatile tool that acts as a wrapper around RAPL, enhancing ease of use.
        \item \textbf{Turbostat:} For assessing CPU utilization or indirect power measurements, contributing to the toolkit available for power analysis.
        \item \textbf{NVIDIA SMI:} NVIDIA System Management Interface (NVIDIA-SMI) is a command-line utility that provides detailed monitoring and management of NVIDIA GPU devices. It allows users to track metrics such as GPU utilization, temperature, and power consumption.
        \item \textbf{Intel PCM:} Intel Performance Counter Monitor (\href{https://www.intel.com/content/www/us/en/developer/articles/tool/performance-counter-monitor.html}{Intel PCM}) is a comprehensive toolset that provides real-time monitoring and analysis of hardware performance counters in Intel processors. It helps users track metrics such as CPU utilization, memory bandwidth, and energy consumption.
    \end{itemize}
    
    \item Software tools that incorporate energy reporting alongside other capabilities:
    \begin{itemize}
        \item \textbf{TensorFlow Profiler:} TensorFlow Profiler is a comprehensive performance analysis tool within TensorFlow that helps users optimize the computational efficiency of their machine learning models. It includes features for monitoring various performance metrics, including energy consumption
        \item \textbf{PyTorch Profiler:} PyTorch Profiler is a performance analysis tool integrated into PyTorch, designed to monitor and optimize the computational efficiency of machine learning models. It includes energy reporting capabilities that allow users to track and reduce the power consumption of their models
        \item \textbf{Vtune:} \href{https://www.intel.com/content/www/us/en/developer/tools/oneapi/vtune-profiler.html}{Intel VTune Profiler} is a performance profiling tool designed to analyze the performance of applications running on Intel processors. It provides detailed insights into CPU, GPU, and memory usage. 
        \item \textbf{Dynatrace:} Dynatrace is an advanced software intelligence platform that offers comprehensive monitoring and analytics for application performance, infrastructure, and user experience.
        \item \textbf{GEOPM}: The Global Extensible Open Power Manager (GEOPM) provides a framework to explore power and energy optimizations on platforms with heterogeneous mixes of computing hardware. Users can monitor their component, node, system, and job energy and power consumption, while safely optimizing system hardware settings to achieve energy efficiency and/or performance objectives. 

    \end{itemize}

    \item Sustainability-related tools also include, or rely on these power/energy measurement tools, including:
    \begin{itemize}
        \item \textbf{Code Carbon:} \href{https://arxiv.org/pdf/2311.10267.pdf#page=6&zoom=100,401,318}{Code Carbon} is often used in the machine learning community for estimating both power and emissions of larger workloads like ML training runs. One publication did both physical measurement and Code Carbon measurement and estimated that the difference between the two on the hardware used is E (kWh) = 1.059 · codecarbon kWh.        
        
        \item \textbf{ML CO${_2}$ Impact:} \href{https://mlco2.github.io/impact/}{ML CO${_2}$ Impact} is post-hoc estimating tool. It calculates the carbon footprint of machine learning models based on recorded resource usage and computational data by by using user-provided input in the \href{https://calculator.linkeddata.es/}{calculator}.

        \item \textbf{Schaphandre:} An open-source tool designed to measure and monitor the energy consumption of servers and data centers, using RAPL (Running Average Power Limit) for power data collection and providing insights into energy efficiency.

        \item \textbf{Green Algorithms:} \href{https://www.green-algorithms.org/}{Green Algos} is an online calculator that uses power estimates and average emission levels to create approximations of emissions for larger jobs, plus code to integrate calculations across compute resources using SLURM.

        \item \textbf{Datavizta:} \href{https://dataviz.boavizta.org/serversimpact}{Datavizta} estimates of a variety of environmental impacts of a server, device, or cloud instance across the hardware’s lifespan, including the environmental cost of manufacturing.

        \item \textbf{Cloud Carbon Footprint:} It works with cloud provider billing to estimate power and emissions based on \href{https://www.etsy.com/codeascraft/cloud-jewels-estimating-kwh-in-the-cloud/}{Etsy Cloud Jewels} on a monthly basis across a whole cloud instance (i.e., not per job). Their \href{https://www.cloudcarbonfootprint.org/docs/methodology}{methodology} includes estimating the embodied and operational emissions.

        \item \textbf{PowerAPI:} \href{https://powerapi.org/getting_started/}{PowerAPI} provides energy usage data by integrating with hardware sensors and power meters to collect real-time power consumption measurements. It then aggregates and analyzes this data to offer detailed insights into the energy usage of specific components or applications.

        \item \textbf{Kepler:} Kubernetes Efficient Power Level Exporter (\href{https://sustainable-computing.io/}{Kepler}) is a tool designed to monitor and report the power consumption of Kubernetes clusters. It integrates with Kubernetes to provide detailed insights into the energy usage of containerized applications.

        \item \textbf{Impact Framework:} Green Software Foundation \href{https://if.greensoftware.foundation/}{Impact Framework} creates a standardized manifest for sharing information about your software, including energy

        \item \textbf{EcoCI:} A GitHub Action developed by Green Coding that estimates the energy in Joules of a CI run that runs on Microsoft Azure VMs.

        \item \textbf{AWS Carbon Footprint Tool:} Offered by Amazon Web Services, the \href{https://aws.amazon.com/aws-cost-management/aws-customer-carbon-footprint-tool/}{AWS Carbon Footprint Tool} helps measure and manage the carbon footprint of AWS services and resources.        
        
        \item \textbf{Green Software Repositories:} Two Running list of green software tools, \href{https://github.com/topics/green-software}{here} and \href{https://github.com/stars/davidkopp/lists/green-software}{here}, that include tools adapted for measuring specific software types(generative AI inferences, continuous integration (CI) tasks, etc.).

    \end{itemize}

    \item Power Benchmarking:
    \begin{itemize}
        \item \textbf{SPECPower:} SPECPower is a benchmarking tool designed to measure and evaluate the power and performance efficiency of server-class computers from active idle to 100\% utilization. They also publish annual performance \href{https://www.spec.org/power_ssj2008/results/}{results} of various servers.

        \item \textbf{SERT:} The Server Energy Rating Tool (SERT) is a benchmarking tool developed by the U.S. Environmental Protection Agency (EPA) that measures and evaluates the energy efficiency of server hardware.

        \item \textbf{Turbostress} Turbostress is a benchmarking tool designed to assess the power consumption and performance of processors under stress conditions. It provides insights into how CPUs handle high workloads and power demands.
    \end{itemize}
\end{enumerate}

\cleardoublepage
\label{sec:Bib}
\nocite{*} 
\printbibliography[title={\LARGE References}]
\addcontentsline{toc}{chapter}{References}
\end{document}